\documentclass[10pt,aps,prx,twocolumn,notitlepage,showpacs,superscriptaddress]{revtex4-1}
\usepackage{amsthm,amsmath,amsfonts,amssymb,verbatim,color}
\usepackage{graphicx}
\usepackage{subfigure}
\usepackage{bm}
\usepackage{epsfig,slashed}
\usepackage[T1]{fontenc}
\usepackage[colorlinks=true,citecolor=blue,linkcolor=blue,urlcolor=blue]{hyperref}
\usepackage{psfrag}
\begin{document}

\newcommand{\tr}{\mathop{\mathrm{Tr}}}
\newcommand{\bsigma}{\boldsymbol{\sigma}}
\newcommand{\re}{\mathop{\mathrm{Re}}}
\newcommand{\im}{\mathop{\mathrm{Im}}}
\renewcommand{\b}[1]{{\boldsymbol{#1}}}
\newcommand{\diag}{\mathrm{diag}}
\newcommand{\sign}{\mathrm{sign}}
\newcommand{\sgn}{\mathop{\mathrm{sgn}}}
\renewcommand{\c}[1]{\mathcal{#1}}
\renewcommand{\d}{\text{\dj}}

\renewcommand{\vec}[1]{{\boldsymbol{#1}}}

\newcommand{\mb}{\bm}
\newcommand{\ua}{\uparrow}
\newcommand{\da}{\downarrow}
\newcommand{\ra}{\rightarrow}
\newcommand{\la}{\leftarrow}
\newcommand{\mc}{\mathcal}
\newcommand{\bs}{\boldsymbol}
\newcommand{\lra}{\leftrightarrow}
\newcommand{\nn}{\nonumber}
\newcommand{\half}{{\textstyle{\frac{1}{2}}}}
\newcommand{\mf}{\mathfrak}
\newcommand{\MF}{\text{MF}}
\newcommand{\IR}{\text{IR}}
\newcommand{\UV}{\text{UV}}
\newcommand{\be}{\begin{equation}}
\newcommand{\ee}{\end{equation}}

\DeclareGraphicsExtensions{.png}

\title{Rashba scattering in the low-energy limit}

\author{Joel Hutchinson}
\email[electronic address: ]{jhutchin@ualberta.ca}
\affiliation{Department of Physics, University of Alberta, Edmonton, Alberta T6G 2E1, Canada}
\affiliation{Theoretical Physics Institute, University of Alberta, Edmonton, Alberta T6G 2E1, Canada}

\author{Joseph Maciejko}
\email[electronic address: ]{maciejko@ualberta.ca}
\affiliation{Department of Physics, University of Alberta, Edmonton, Alberta T6G 2E1, Canada}
\affiliation{Theoretical Physics Institute, University of Alberta, Edmonton, Alberta T6G 2E1, Canada}
\affiliation{Canadian Institute for Advanced Research, Toronto, Ontario M5G 1Z8, Canada}

\date\today

\begin{abstract}
We study potential scattering in a two-dimensional electron gas with Rashba spin-orbit coupling in the limit that the energy of the scattering electron approaches the bottom of the lower spin-split band. Focusing on two spin-independent circularly symmetric potentials, an infinite barrier and a delta-function shell, we show that scattering in this limit is qualitatively different from both scattering in the higher spin-split band and scattering of electrons without spin-orbit coupling. The scattering matrix is purely off-diagonal with both off-diagonal elements equal to one, and all angular momentum channels contribute equally; the differential cross section becomes increasingly peaked in the forward and backward scattering directions; the total cross section exhibits quantized plateaus. These features are independent of the details of the scattering potentials, and we conjecture them to be universal. Our results suggest that Rashba scattering in the low-energy limit becomes effectively one-dimensional.
\end{abstract}


\maketitle

\section{Introduction}

In crystalline solids with time-reversal and inversion symmetries, electronic energy bands are doubly degenerate. If inversion symmetry is broken by the crystal structure or by electric fields (internal or externally applied), spin-orbit coupling generically leads to a splitting of the bands. In two-dimensional (2D) electron gases where inversion symmetry is broken for structural reasons~\cite{winkler}, e.g., by band bending at the surface of a 3D solid or by an asymmetric confinement potential in semiconductor quantum wells, this effect is most simply described by the Rashba model~\cite{vasko1979,bychkov1984}. In this model [Eq.~(\ref{RashbaH})], the usual Hamiltonian $H_0=\b{k}^2/2m$ for an electron with momentum $\b{k}$ and effective mass $m$ is augmented by a term $H_R=\lambda\hat{\b{z}}\cdot(\bsigma\times\b{k})$ linear in $\b{k}$ and explicitly dependent on the electron spin $\bsigma$, where $\hat{\b{z}}$ is a unit vector normal to the plane of the 2D electron gas and $\lambda$ is the Rashba coupling, with units of velocity. This extra term can be interpreted either as a spin-dependent vector potential or a momentum-dependent Zeeman field, which suggests the possibility of manipulating the electron spin by electric means. Building on this idea, the seminal Datta-Das spin transistor proposal~\cite{datta1990} launched an intense investigation of Rashba systems as promising material platforms for spintronic devices~\cite{zutic2004} that continues to this day~\cite{manchon2015}.

By contrast with conventional 2D electron gases without spin-orbit coupling, Rashba systems are characterized by two qualitatively distinct energy regimes (Fig.~\ref{fig:dispersion}): the positive-energy regime $E>0$ and the negative-energy regime $E<0$, separated by a Dirac point at $\b{k}=0$, $E=0$. While both regimes are characterized by two spin-split Fermi surfaces, in the $E>0$ regime the density of states is constant, as in conventional 2D electron gases, while in the $E<0$ regime it displays an inverse square-root singularity at the band bottom $E=-E_0$, with $E_0=m\lambda^2/2$ (see, e.g., Fig.~1(c) in Ref.~\cite{cappelluti2007}). This divergence is a consequence of the fact that the band bottom in Rashba systems is a degenerate ring of states with momentum $|\b{k}|=k_0$ where $k_0=m\lambda$, rather than a single point $\b{k}=0$ as for a conventional parabolic dispersion. The divergent density of states leads to an increased phase space for scattering at low energies, and is known to enhance various symmetry-breaking instabilities in the presence of attractive~\cite{cappelluti2007,takei2012} or repulsive~\cite{berg2012,silvestrov2014,ruhman2014,bahri2015} two-body interactions. The discovery of materials with extremely large Rashba splittings such as the polar semiconductor BiTeI with $E_0\approx 100$~meV~\cite{ishizaka2011}, a Bi-trimer adlayer on the Si(111) surface with $E_0\approx 140$~meV~\cite{gierz2009}, and the Bi/Ag(111) surface alloy with $E_0\approx 200$~meV~\cite{ast2007} suggests that the $E<0$ regime is experimentally accessible. The recent demonstration of synthetic spin-orbit coupling in cold atomic gases~\cite{galitski2013} may lead to further possibilities.

In this paper we explore the single-particle scattering of Rashba electrons off circularly symmetric, finite-range potentials in the negative-energy regime $E<0$, with a focus on the low-energy limit $E\rightarrow-E_0$. While potential scattering in Rashba systems has been studied before in the $E>0$ regime~\cite{cserti2004,Yeh2006,walls2006,palyi2006,csordas2006,palyi2007}, little attention has been payed to the $E<0$ regime in this context. We find peculiar features in the low-energy limit: (1) the $S$-matrix for a partial wave of angular momentum $l$ approaches a purely off-diagonal form with both off-diagonal elements equal to one, independent of $l$ [Eq.~(\ref{UniversalSmatrix})]; (2) the differential cross section becomes quasi-1D, with only forward and backward scattering allowed [Eq.~(\ref{eq:diffxsec_E0})]; (3) the total cross section exhibits quantized plateaus [Eq.~(\ref{quantizedsigma})]. Remarkably, these results hold for both the infinite barrier (Sec.~\ref{sec:harddisk}) and infinitely thin shell (Sec.~\ref{sec:hardshell}) potentials considered here, with no dependence on the details of the potentials such as range and amplitude. These features contrast severely with both $E>0$ scattering in the Rashba case and low-energy scattering in the conventional case without spin-orbit coupling; we conjecture they are universal properties of Rashba scattering in the low-energy limit. Our results suggest that impurity scattering in low-density Rashba systems where the Fermi energy $E_F$ is much less than the Rashba splitting $E_0$ should be qualitatively different from the high-density regime.

\section{Rashba spin-orbit coupling}

\begin{figure}[t]
	\centering
	\psfrag{l}{$\sin(x)$}
	\includegraphics[width=\columnwidth]{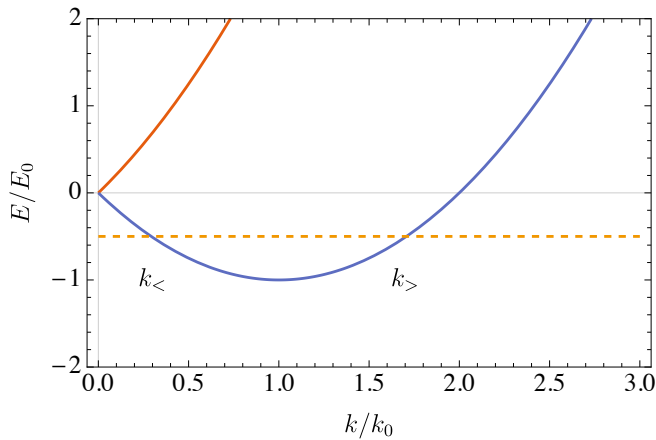}
\caption{2D single-particle Rashba dispersion, with energy $E$ in units of $E_0=m\lambda^2/2$ and modulus of the wave vector $k$ in units of $k_0=m\lambda$, where $m$ is the effective mass of the electron and $\lambda$ is the Rashba coupling. For positive energies, there are two helicity bands of electron states. At negative energies (dashed line), there is a single helicity band, but one ring of states with wave vector magnitude $k_>$, and one with wave vector magnitude $k_<$ distinguished by the sign of the group velocity $\b{v}_g\cdot\hat{\b{k}}=(k-k_0)/m$.}
\label{fig:dispersion}
\end{figure}

We begin with the single-particle Rashba Hamiltonian in two dimensions~\cite{bychkov1984},
\be\label{RashbaH}
H(\b{k})=\frac{\b{k}^2}{2m}+\lambda \hat{\b{z}}\cdot(\b{\sigma}\times\b{k}),
\ee
where $\b{k}=(k_x,k_y)$ is the electron wave vector, $\bsigma=(\sigma_x,\sigma_y)$ is a vector of Pauli matrices, $m$ is the effective mass of the electron, and $\lambda$ is the Rashba coupling, with units of velocity (we work in units such that $\hbar=1$). Diagonalization gives a spin-split spectrum
\be
E_\pm(k)=\frac{k^2}{2m}\pm\lambda k,
\ee
which has a ring of degenerate points for each wave vector magnitude $k$. The spin vector $\langle \b{S}\rangle=\frac{1}{2}\langle\bsigma\rangle$ is locked orthogonally to $\b{k}$, but in opposite directions for the $E_+$ and $E_-$ bands, which we refer to as the positive- and negative-helicity bands respectively. The two spin-split paraboloids have minima at $k_0\equiv m\lambda$, giving a band-bottom energy of $-E_0=-m\lambda^2/2$. We are interested in the low-energy regime ($E$ near $-E_0$) in which $E$ is strictly negative. In this regime only the negative-helicity band is accessible, however, there are still two degenerate rings at any given energy with wave vector magnitudes
\begin{eqnarray}
k_>&=&k_0+\sqrt{2m(E_0-|E|)}\\
k_<&=&k_0-\sqrt{2m(E_0-|E|)},
\end{eqnarray}
as indicated in Fig.~\ref{fig:dispersion}. Throughout this paper, we measure wave vectors/inverse lengths in units of $k_0$ and energies in units of $E_0$, so that 
\be\label{eq:kpm}
\frac{k_{\gtrless}}{k_0}=1\pm\sqrt{1-\frac{|E|}{E_0}}.
\ee
By contrast with an ordinary 2D electron gas without spin-orbit coupling, the Hamiltonian contains a length scale $\sim k_0^{-1}$ at the band bottom in the absence of a scattering potential.

To solve the scattering problem, we require the Hamiltonian in position-space polar coordinates $(r,\theta)$:
\be\label{Hpolar}
H=\begin{pmatrix}
-\frac{1}{2m}(\partial_r^2+\frac{1}{r}\partial_r+\frac{1}{r^2}\partial_\theta^2) & \lambda e^{-i\theta}(\partial_r-\frac{i}{r}\partial_\theta)\\
-\lambda e^{i\theta}(\partial_r+\frac{i}{r}\partial_\theta)& -\frac{1}{2m}(\partial_r^2+\frac{1}{r}\partial_r+\frac{1}{r^2}\partial_\theta^2) \\
\end{pmatrix},
\ee
whose eigenfunctions can be expanded in partial waves as
\be
\psi(r,\theta)=\sum_{l=-\infty}^\infty e^{il\theta}
\begin{pmatrix}
R_l(r)\\
e^{i\theta} R_{l+1}(r)\\
\end{pmatrix}.
\ee
The radial functions $R_l(r)$ are linear combinations of incoming and outgoing Hankel functions $H^\pm_l(kr)$, defined as $H_l^\pm(x)=J_l(x)\pm iN_l(x)$ where $J_l(x)$ and $N_l(x)$ are Bessel functions of the first and second kind (Neumann functions), respectively. We consider elastic scattering at negative energy $E$, so that $k$ in the argument of the Hankel functions can take on either value $k_\gtrless$ satisfying \eqref{eq:kpm}. There are four independent solutions to the Schr{\"o}dinger equation, and a generic eigenfunction $\Psi$ at energy $E$ for the free-particle problem may be written as
\begin{widetext}
\begin{equation}
\label{eq:psi_generic}
\Psi(r,\theta)=\sum_{l=-\infty}^\infty e^{il\theta}
\left[ a_l\begin{pmatrix} H_l^+(k_<r)\\ -H_{l+1}^+(k_<r)e^{i\theta}\end{pmatrix} +b_l\begin{pmatrix} H_l^-(k_<r)\\ -H_{l+1}^-(k_<r)e^{i\theta}\end{pmatrix}
+c_l\begin{pmatrix} H_l^+(k_>r)\\ -H_{l+1}^+(k_>r)e^{i\theta}\end{pmatrix}+d_l\begin{pmatrix} H_l^-(k_>r)\\ -H_{l+1}^-(k_>r)e^{i\theta}\end{pmatrix}\right],
\end{equation}
\end{widetext}
where $a_l$, $b_l$, $c_l$, and $d_l$ are arbitrary coefficients.


\section{Hard-disk scattering}\label{sec:harddisk}

We now add to the free-particle Hamiltonian (\ref{RashbaH}) a scattering potential $V$. We first consider single-electron scattering off an infinite circular barrier
\be\label{Vharddisk}
V=\begin{cases} 
\infty, & r\leq R,\\
0, & r>R.
\end{cases}
\ee
Because the potential vanishes identically for $r>R$, eigenstates of the full Hamiltonian with energy $E$ obey the free-particle expansion (\ref{eq:psi_generic}) in that region. In that region, the wave function consists of an incident plane wave $\psi^\textrm{in}_\gtrless$ with definite wave vector $k_\gtrless\hat{\b{x}}$, as well as outgoing scattered waves with each of the allowed wave vectors. In a typical scattering problem, the outgoing states consist of $H^+(kr)$ radial functions, which combines with the fact that the group velocity $\b{v}_g$ points in the same direction as the wave vector $\b{k}$ to ensure that the probability current carried by an outgoing state is directed radially outwards. However, in the Rashba problem the expectation value of the group velocity $\b{v}_g=\nabla_\b{k}H_0(\b{k})$ in states of negative helicity is $\langle\b{v}_g\rangle=(k-k_0)\hat{\b{k}}/m$. For energies below the Dirac point, the $k_<$ states have group velocity antiparallel to the wave vector, thus the outgoing $k_<$ states should be accompanied by $H^-(kr)$ radial functions to carry a probability current directed radially outwards. For an incident wave in the $k_>$ state, the wave function for $r>R$ can be written as
\begin{equation}
\psi_>(r,\theta)=\psi^{\rm in}_>(r,\theta)+\sum_{l=-\infty}^\infty e^{il\theta} [\psi^l_c(r,\theta)+\psi^l_b(r,\theta)],\label{eq:wavefunck+}
\end{equation}
where 
\begin{eqnarray}
\psi^l_c(r,\theta)&\equiv& \left(c_l-\frac{i^l}{2\sqrt{2}}\right)\begin{pmatrix} H_l^+(k_>r)\\ -H_{l+1}^+(k_>r)e^{i\theta}\end{pmatrix},\\ 
\psi^l_b(r,\theta)&\equiv& b_l\begin{pmatrix} H_l^-(k_<r)\\ -H_{l+1}^-(k_<r)e^{i\theta}\end{pmatrix},
\end{eqnarray}
while for an incident wave in the $k_<$ state, we have
\begin{equation}
\psi_<(r,\theta)=\psi^{\rm in}_<(r,\theta)+\sum_{l=-\infty}^\infty e^{il\theta}[\psi^l_{\tilde{c}}(r,\theta)+\psi^l_{\tilde{b}}(r,\theta)],\label{eq:wavefunck-}
\end{equation}
where
\begin{eqnarray}
\psi^l_{\tilde{c}_l }(r,\theta)&\equiv&\tilde{c}_l \begin{pmatrix} H_l^+(k_>r)\\ -H_{l+1}^+(k_>r)e^{i\theta}\end{pmatrix},\nonumber\\
\psi^l_{\tilde{b}_l}(r,\theta)&\equiv&\left(\tilde{b}_l-\frac{i^l}{2\sqrt{2}}\right)\begin{pmatrix} H_l^-(k_<r)\\ -H_{l+1}^-(k_<r)e^{i\theta}\end{pmatrix}.
\end{eqnarray}
In these expressions $b_l$, $c_l$, $\tilde{b}_l$, and $\tilde{c}_l$ are coefficients to be determined by a solution of the scattering problem.

\begin{figure}[t]
	\centering
	\includegraphics[width=\columnwidth]{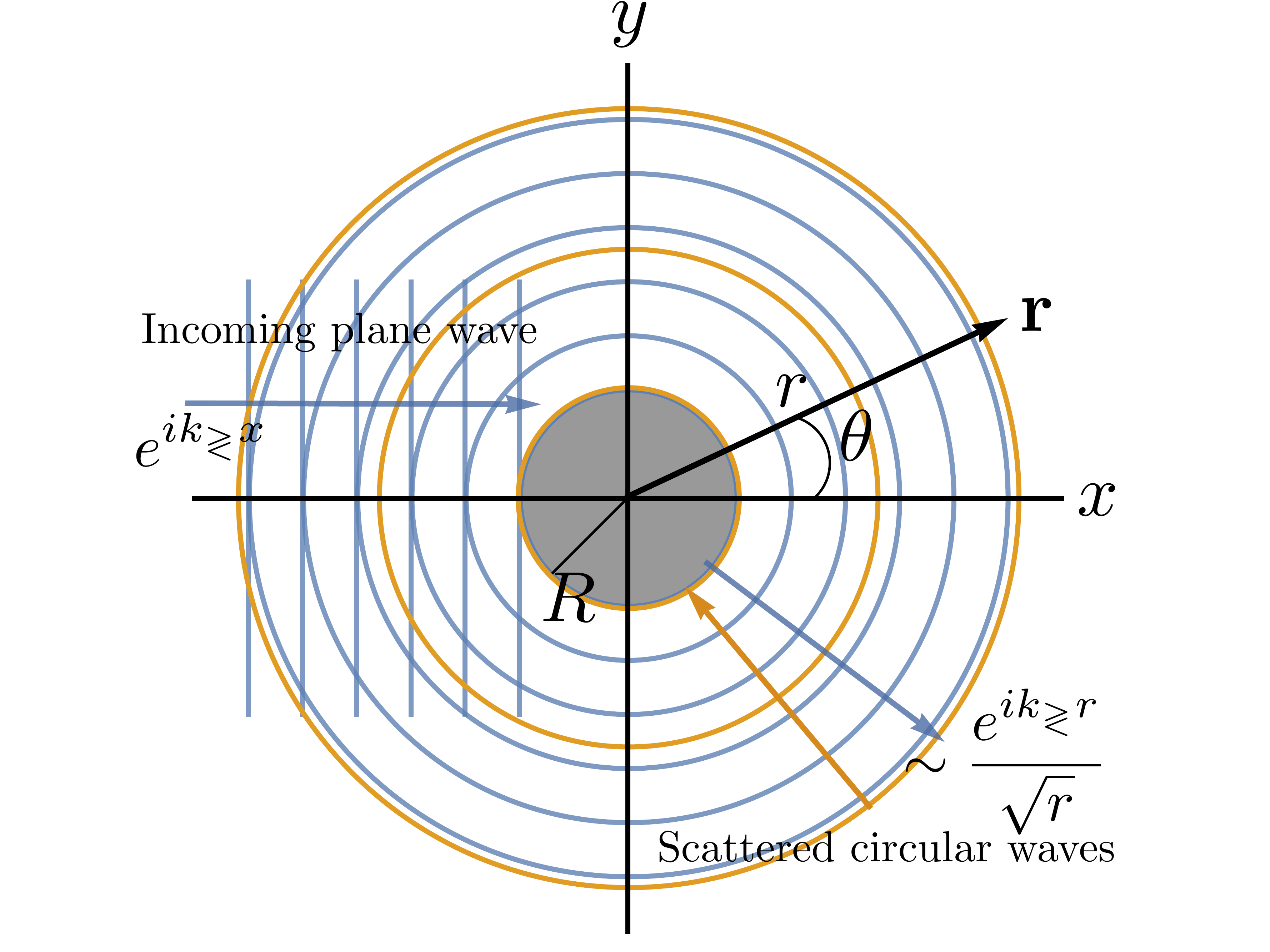}
\caption{Plane wave scattering off an infinite circular barrier. There are two circular scattered states (blue and orange) of different wavelengths corresponding to the $k_>$ and $k_<$ states, respectively.}
\label{fig:scatter}
\end{figure}
The incident plane wave can itself be decomposed into partial waves:
\begin{eqnarray}
\psi^{\rm in}_\gtrless(r,\theta)&=&\frac{1}{\sqrt{2}}\begin{pmatrix}1\\ i\end{pmatrix}e^{ik_\gtrless x}\nonumber\\
&=&\sum_{l=-\infty}^\infty\frac{i^l}{2\sqrt{2}}e^{il\theta}\bigg[\begin{pmatrix} H_l^+(k_\gtrless r)\\ -H_{l+1}^+(k_\gtrless r)e^{i\theta}\end{pmatrix}\nonumber\\
&&+\begin{pmatrix} H_l^-(k_\gtrless r)\\ -H_{l+1}^-(k_\gtrless r)e^{i\theta}\end{pmatrix}\bigg].
\end{eqnarray}
The infinite potential barrier (\ref{Vharddisk}) forces the wave function to vanish at $r=R$,
\be\label{eq:match}
\psi_\gtrless(R,\theta)=\begin{pmatrix} 0 \\ 0\end{pmatrix}.
\ee
Imposing this condition in Eq.~\eqref{eq:wavefunck+} and \eqref{eq:wavefunck-} gives four equations from which we obtain the unknown coefficients $b_l$, $\tilde{b}_l$, $c_l$, $\tilde{c}_l$:
\begin{eqnarray}
b_l&=&\frac{1}{\Delta_l}\bigg(H^+_l(k_>R)H^-_{l+1}(k_>R)-H^-_l(k_>R)H^+_{l+1}(k_>R)\bigg),\nonumber\\
c_l&=&\frac{1}{\Delta_l}\bigg(H^-_l(k_>R)H^-_{l+1}(k_<R)-H^-_l(k_<R)H^-_{l+1}(k_>R)\bigg),\nonumber\\
\tilde{b}_l&=&\frac{1}{\Delta_l}\bigg(H^+_l(k_>R)H^+_{l+1}(k_<R)-H^+_l(k_<R)H^+_{l+1}(k_>R)\bigg),\nonumber\\
\tilde{c}_l &=&\frac{1}{\Delta_l}\bigg(H^+_l(k_<R)H^-_{l+1}(k_<R)-H^-_l(k_<R)H^+_{l+1}(k_<R)\bigg),\nonumber
\end{eqnarray}
where we have defined 
\be
\Delta_l\equiv\frac{2\sqrt{2}}{i^l}\bigg(H^-_l(k_<R)H^+_{l+1}(k_>R)-H^+_l(k_>R)H^-_{l+1}(k_<R)\bigg).
\ee

\subsection{$S$-matrix}

The four coefficients above determine the $S$-matrix for this scattering problem. The $S$-matrix is the unitary transformation that connects asymptotic states in the incoming circular basis ($e^{\mp i(k_\gtrless r-l\pi/2)}/\sqrt{r}$) to asymptotic states in the outgoing circular basis ($e^{\pm i(k_\gtrless r-l\pi/2)}/\sqrt{r}$).  Using the asymptotic form of the Hankel functions for large argument $x\gg 1$,
\be\label{eq:HankelAsymp}
H_l^\pm(x)\approx\sqrt{\frac{2}{\pi x}}e^{\pm i(x-l\pi/2-\pi/4)},
\ee
we obtain the $S$-matrix in angular momentum channel $l$,
\be\label{Smatrix}
S^l=\begin{pmatrix}
S_{>>} & S_{><}\\
S_{<>} & S_{<<}
\end{pmatrix}=
\frac{2\sqrt{2}}{i^l}\begin{pmatrix} c_l & b_l\sqrt{\frac{k_>}{k_<}}\\
\tilde{c}_l \sqrt{\frac{k_<}{k_>}} & \tilde{b}_l\end{pmatrix}.
\ee
Using the explicit expressions given earlier for the coefficients $b_l$, $\tilde{b}_l$, $c_l$, $\tilde{c}_l$, as well as the Wronskian identity
\be
H^+_l(z)H^-_{l+1}(z)-H^+_{l+1}(z)H^-_l(z)=\frac{4i}{\pi z},
\ee
we find that $S_{><}^l=S_{<>}^l$; this is a consequence of time-reversal symmetry combined with reflection symmetry about the $x$ axis (see Appendix~\ref{app:symmetry}). In this case, there are two independent unitarity conditions on the $S$-matrix,
\begin{equation}\label{unitcond}
|c_l|^2+|b_l|^2\frac{k_>}{k_<}=\frac{1}{8},\hspace{5mm}
b_l\tilde{b}_l^*=-c_lb_l^*,
\end{equation}
which are satisfied by the coefficients given above.
Unitarity of the $S$-matrix should be equivalent to the continuity equation:
\be\label{eq:continuity}
\int_A d^2r\,\nabla\cdot\vec{j}(\vec{r},t)=\int_0^{2\pi}d\theta\,\vec{r}\cdot\vec{j}(\b{r},\theta) = 0.
\ee
The flux current density is readily found for the Rashba system to be $\vec{j}(\vec{r},t)=\vec{j}_{K}+\vec{j}_R$, where the kinetic and Rashba current densities are $\vec{j}_K\equiv-\frac{i}{2m}(\psi^\dagger\nabla\psi-\nabla\psi^\dagger\psi)$, and $\vec{j}_R\equiv-\lambda\psi^\dagger(\vec{\sigma}\times\hat{\b{z}})\psi$ respectively. By angular momentum conservation, we can replace $\psi$ in the above definitions with its partial wave component. 
The integral over the ring in the continuity equation (\ref{eq:continuity}) is then evaluated for each partial wave. For an incident $k_>$ wave, one obtains
\begin{eqnarray}
\int^{2\pi}_0 d\theta\,\vec{r}\cdot \vec{j}_{K}&=&-\frac{16\lambda}{\sqrt{k_>k_<}}b_l^*d_l\cos[(k_<-k_>)r]\nonumber\\
&&+\frac{8}{m}(-|b_l|^2+|c_l|^2-1/8)\label{eq:fluxK},\\
\int^{2\pi}_0 d\theta\,\vec{r}\cdot \vec{j}_{R}&=&\frac{16\lambda}{\sqrt{k_>k_<}}b_l^*d_l\cos[(k_<-k_>)r]\nonumber\\
&&+8\lambda\bigg(\frac{|b_l|^2}{k_<}-\frac{|c_l|^2}{k_>}+\frac{1/8}{k_>}\bigg).\label{eq:fluxR}
\end{eqnarray}
The first term in each equation is an interference term between scattered partial wave components of different wave vectors ($k_>$ and $k_<$). Combining the kinetic and Rashba pieces, we see that the interference terms completely cancel giving
\be
\int^{2\pi}_0 d\theta\, \vec{r}\cdot \vec{j}=\sqrt{2mE+(m\lambda)^2}\bigg(\frac{|b_l|^2}{k_<}+\frac{|c_l|^2}{k_>}-\frac{1/8}{k_>}\bigg),
\ee
so the continuity equation is satisfied by the first unitarity condition in Eq.~(\ref{unitcond}). Repeating the above calculation for an incident $k_<$ wave, gives a second continuity equation:
\be
|\tilde{b}_l|^2+|\tilde{c}_l |^2\frac{k_<}{k_>}=\frac{1}{8},
\ee
which may alternatively be obtained using $S^l_{><}=S^l_{<>}$ in combination with the unitarity conditions \eqref{unitcond}.

\begin{figure}[t]
	\centering
	\includegraphics[width=\columnwidth]{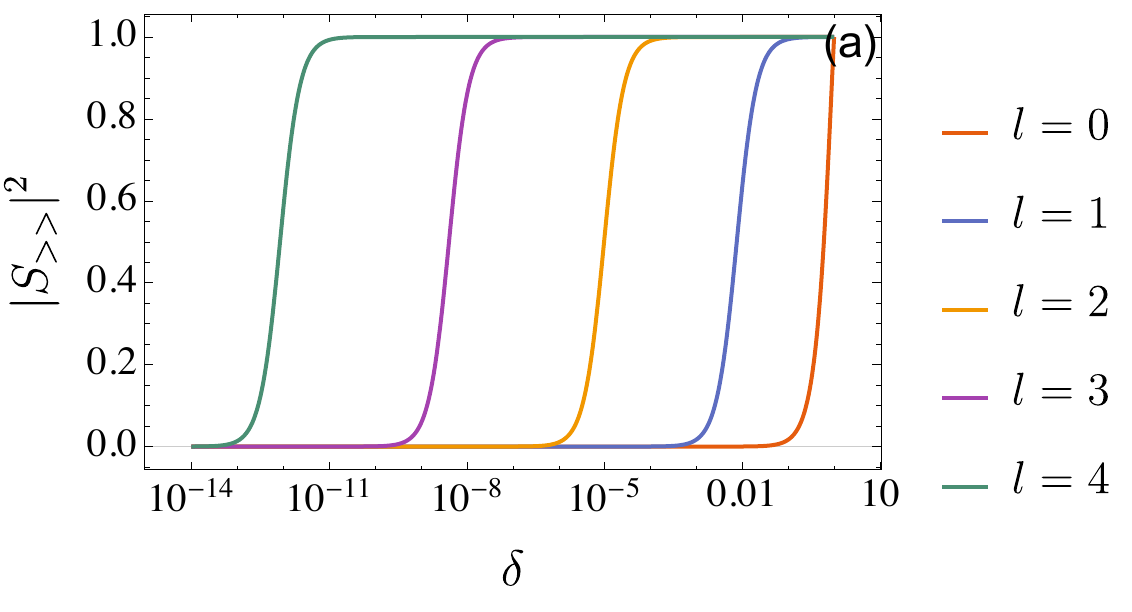}
	\includegraphics[width=\columnwidth]{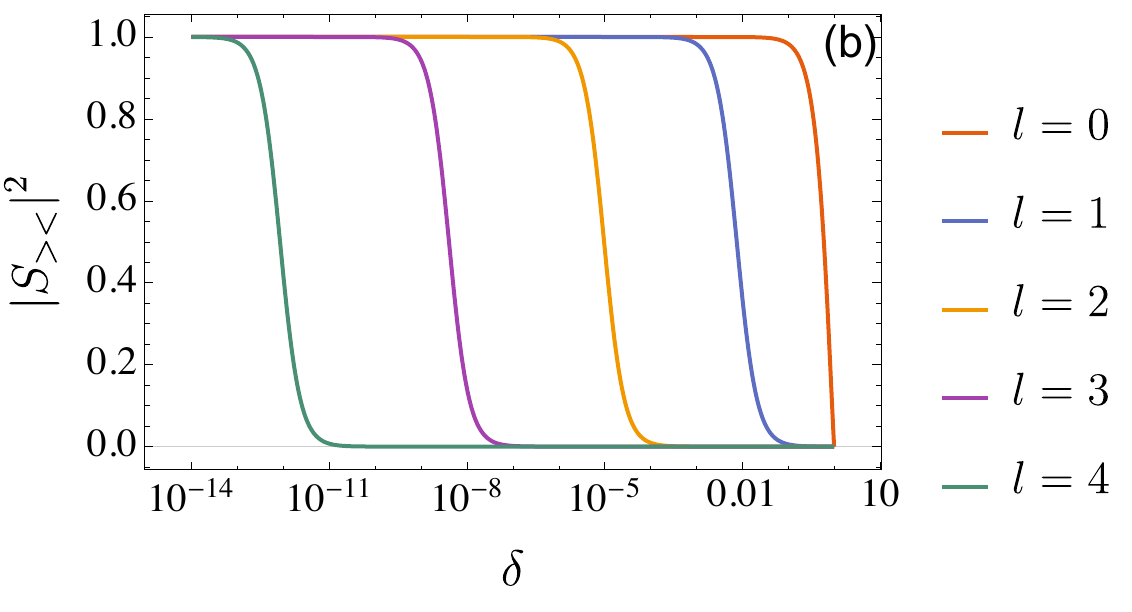}
\caption{(a) Diagonal and (b) off-diagonal transition probabilities from the $S$-matrix elements for partial waves $l=0,1,2,3,4$, as a function of $\delta=\sqrt{1-|E|/E_0}$. In both plots $k_0R=0.1$.}
\label{fig:S_amp2}
\end{figure}

We may plot transition probabilities from the square modulus of the $S$-matrix elements in Eq.~(\ref{Smatrix}). The second equation in \eqref{unitcond} ensures that $|S_{>>}|^2=|S_{<<}|^2$, and symmetry requires $|S_{><}|^2=|S_{<>}|^2$, hence only $|S_{>>}|^2$ and $|S_{><}|^2$ are plotted in Fig.~\ref{fig:S_amp2}. The curves are plotted on a log-linear scale with $\delta\equiv\sqrt{1-|E|/E_0}$ a dimensionless measure of the departure of the energy from the band bottom at $\delta=0$. As the energy approaches the band bottom, fewer partial waves contribute to the diagonal transition probabilities, while more partial waves contribute to the off-diagonal ones. Exactly at the band bottom $E/E_0=-1$, we have $c_l=\tilde{b}_l=0$ and $b_l=-i^l/2\sqrt{2}$, so that the $S$-matrix becomes
\begin{align}\label{UniversalSmatrix}
S^l=\begin{pmatrix}
0 & -1 \\
-1 & 0
\end{pmatrix},
\end{align}
for all $l$, independent of the radius of the scatterer $R$. Scattering is entirely off-diagonal in this limit, and all angular momentum channels contribute equally.

\subsection{Differential cross section}

The differential cross section is a ratio of scattered to incident flux in a particular incoming ($k_\gtrless$) channel,
\be\label{eq:diffxsec}
\left(\frac{d\sigma}{d\theta}\right)_{\gtrless}=r\frac{|\vec{j}^{\rm sc}_\gtrless|}{|\vec{j}^{\rm in}_{\gtrless}|}.
\ee
Using the asymptotic form of the incident and scattered wave functions, the fluxes are given by
\begin{eqnarray}
|\vec{j}^{\rm sc}_>|&=&\frac{k_>-k_0}{mr}\left(|\Phi_{>>}|^2+|\Phi_{><}|^2\right),\\
|\vec{j}^{\rm sc}_<|&=&\frac{k_>-k_0}{mr}\left(|\Phi_{<>}|^2+|\Phi_{<<}|^2\right),\\
\vec{j}^{\rm in}_\gtrless&=&\pm\frac{1}{m}(k_>-k_0)\hat{x},
\end{eqnarray}
so that
\begin{eqnarray}
\left(\frac{d\sigma}{d\theta}\right)_>&=&|\Phi_{>>}|^2+|\Phi_{><}|^2,\label{eq:diffxsec+}\\
\left(\frac{d\sigma}{d\theta}\right)_<&=&|\Phi_{<>}|^2+|\Phi_{<<}|^2.\label{eq:diffxsec-}
\end{eqnarray}
We define
\begin{eqnarray}
\Phi_{>>}&=&\sqrt{\frac{4}{\pi k_>}}\sum_{l}\bigg(c_l-\frac{i^l}{2\sqrt{2}}\bigg)e^{il(\theta-\pi/2)},\label{eq:phiplus}\\
\Phi_{><}&=&\sqrt{\frac{4}{\pi k_<}}\sum_{l}b_le^{il(\theta+\pi/2)},\label{eq:phiminus}\\
\Phi_{<>}&=&\sqrt{\frac{4}{\pi k_>}}\sum_{l}\tilde{c}_le^{il(\theta-\pi/2)},\\
\Phi_{<<}&=&\sqrt{\frac{4}{\pi k_<}}\sum_{l}\bigg(\tilde{b}_l-\frac{i^l}{2\sqrt{2}}\bigg)e^{il(\theta+\pi/2)},
\end{eqnarray}
where sums over $l$ range from $-\infty$ to $\infty$.


\begin{figure}
	\centering
	\includegraphics[width=\columnwidth]{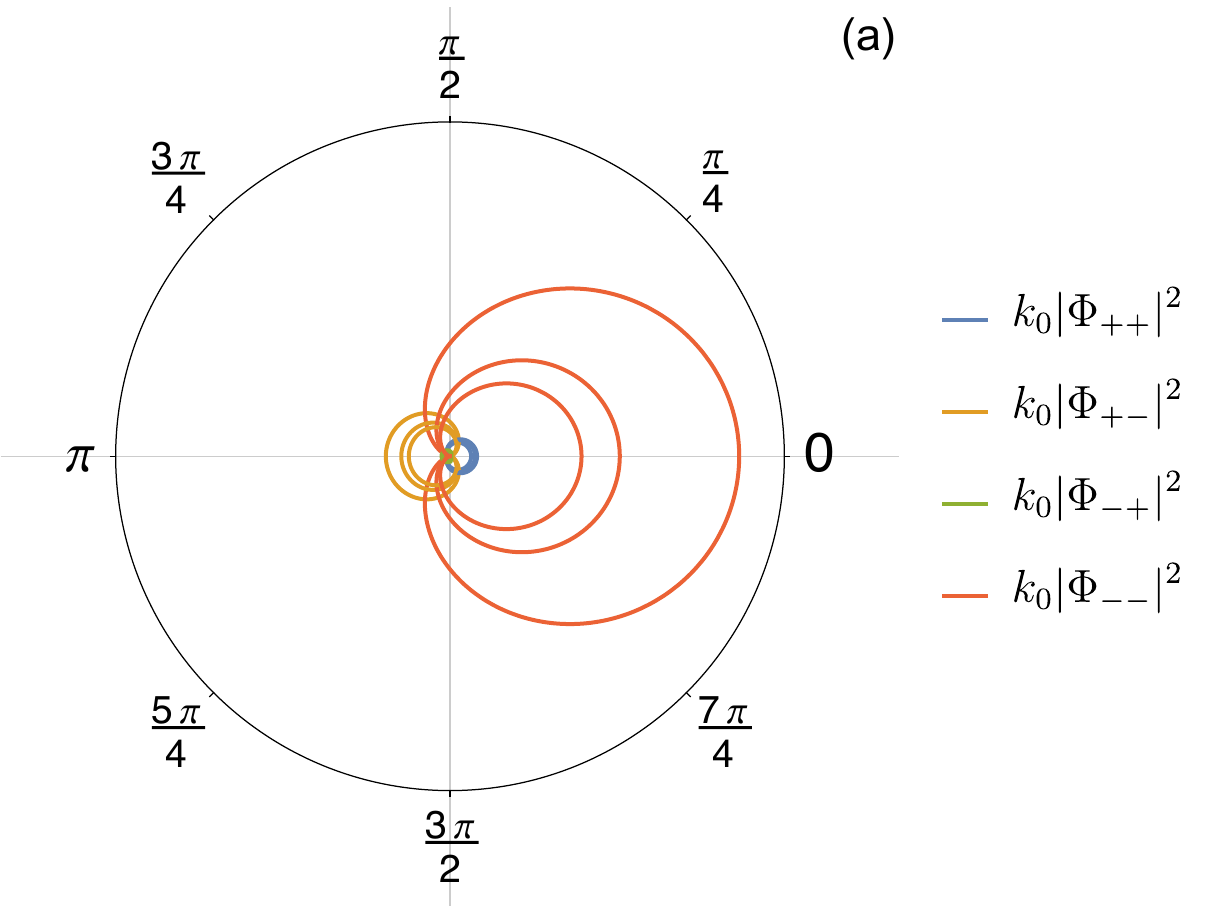}
	\includegraphics[width=\columnwidth]{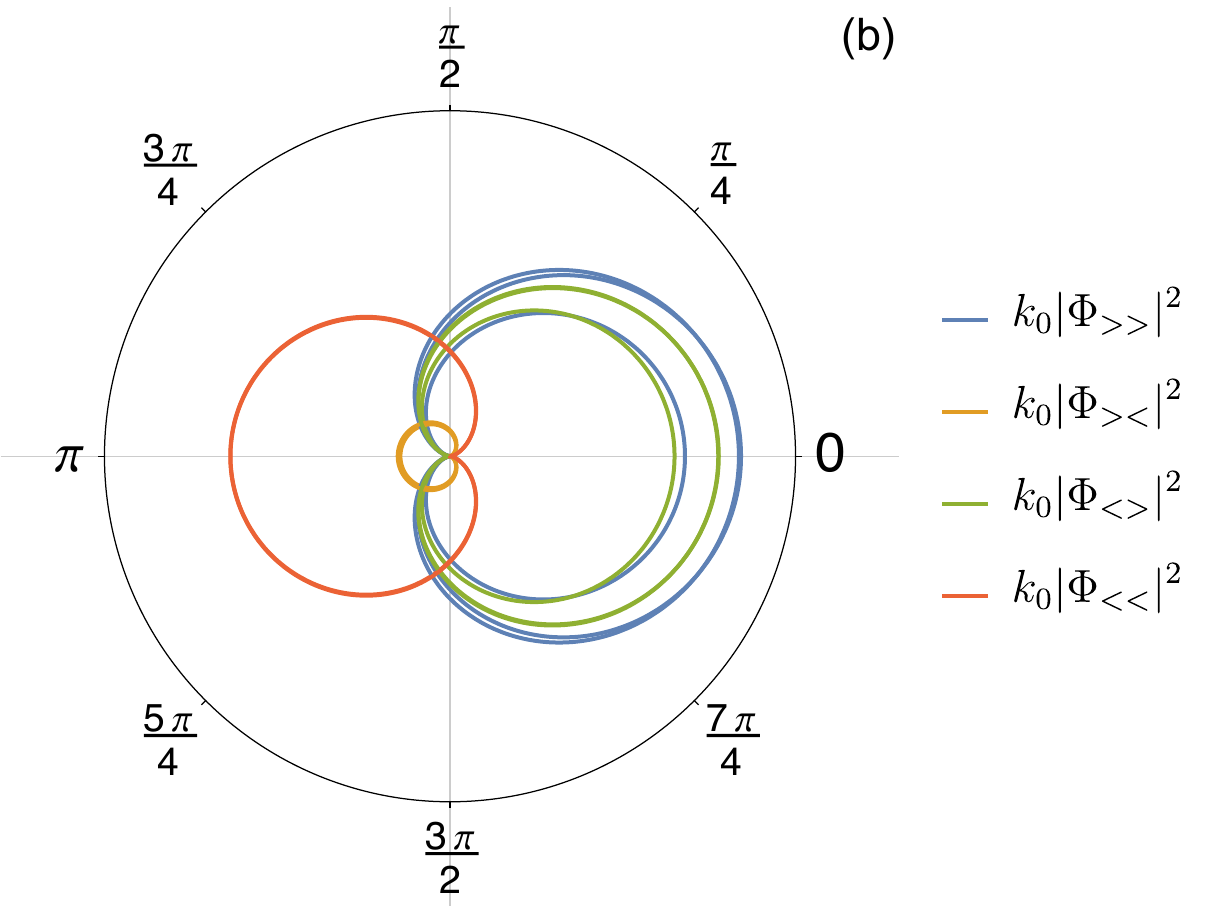}
	\includegraphics[width=\columnwidth]{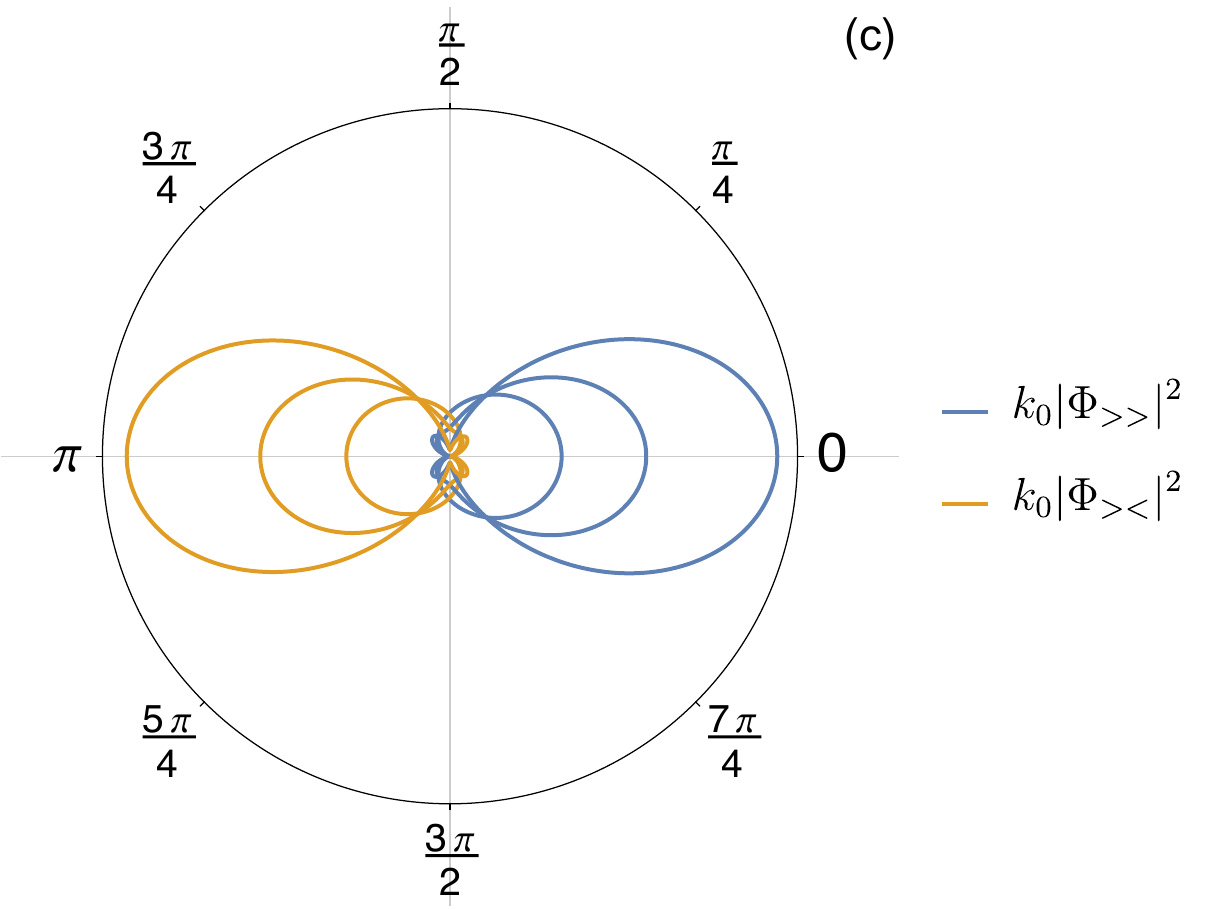}
\caption{Polar plots of differential cross section for scattering between: (a) helicity bands at positive energies ($E=2E_0$, $E=4E_0$, $E=6E_0$), (b) $k_\gtrless$ states at negative energies ($E=-0.01E_0$, $E=-0.5E_0$, $E=-0.99E_0$), and (c) $k_\gtrless$ states near the band bottom  ($E=-0.999E_0$, $E=-0.9999E_0$, $E=-0.99999E_0$).  In each plot, $k_0R$ is set to $0.1$. The radius of each curve is the magnitude of $k_0|\Phi_{ii}|^2$. In the bottom figure, there is no visible distinction between $|\Phi_{><}|^2$ and $|\Phi_{<<}|^2$, as with $|\Phi_{>>}|^2$ and $|\Phi_{<>}|^2$, so only one of each is plotted.}
\label{fig:diffxsec}
\end{figure}

We plot the differential cross section in units of $k_0^{-1}$ in Fig.~\ref{fig:diffxsec}. From panel (c), we see that the differential cross section in the incoming $k_>$ channel ($|\Phi_{>>}|^2+|\Phi_{><}|^2$) becomes increasingly anisotropic with peaks at $\theta=0$ (forward scattering) and $\theta=\pi$ (backscattering) as $E$ tends to the band bottom $-E_0$. Using the observation that in this limit, $c_l=\tilde{b}_l=0$ and $\tilde{c}_l=b_l= -i^l/2\sqrt{2}$, the sums over $l$ in Eq.~(\ref{eq:phiplus}) and (\ref{eq:phiminus}) can be performed analytically and we find that the differential cross section at the band bottom formally becomes
\be\label{eq:diffxsec_E0}
\left(\frac{d\sigma}{d\theta}\right)_\gtrless\bigg|_{E=-E_0}=\frac{2\pi}{k_0}\left[\delta^2(\theta)+\delta^2(\theta-\pi)\right].
\ee
At the band bottom, scattering becomes effectively one-dimensional in that only forward and backward scattering are allowed. No such feature occurs in the $E>0$ regime. The non-integrability of the differential cross section at threshold is a common feature of scattering in two dimensions (see Appendix \ref{App:AppendixA}). Unlike conventional scattering though, the divergence here arises from the contribution of an infinite number of partial waves at the threshold energy. Remarkably, Eq.~\eqref{eq:diffxsec_E0} has no $R$ dependence, and is therefore insensitive to the range of the scattering potential. As shown in Appendix~\ref{App:AppendixA}, this is in contrast with scattering of an electron without spin-orbit coupling where the differential cross section near the band bottom depends explicitly on the radius $R$ of the scatterer. In Sec.~\ref{sec:hardshell} we present further evidence that the details of the impurity potential do not affect this result.

For reference we show in Fig.~\ref{fig:diffxsec}(a) the differential cross section for the $E>0$ regime, which was previously worked out by Yeh \emph{et al.}~\cite{Yeh2006}. In this regime $\pm$ refers to the helicity of the band. The anisotropies in the differential cross section can be understood from the fact that the scattering potential is spin-independent. For example, when starting from an incident positive-helicity state, the electron can only forward scatter into a state of the same helicity, since scattering to the negative-helicity state would flip the spin. Likewise, the electron can only backward scatter into the negative-helicity state, since scattering to the positive-helicity state would flip the spin. This is why the differential cross sections vanish at $\theta=\pi$ for the blue curves, and $\theta=0$ for the orange curves.  The same reasoning can be applied to scattering between $k_\gtrless$ states in the negative-energy regime [Fig.~\ref{fig:diffxsec}(b) and (c)]. Here, an incident $k_>$ electron cannot backscatter to another $k_>$ state without flipping its spin. For scattering from $k_>$ to $k_<$, there is a subtlety to this argument. Because the group velocity in the $k_<$ state is directed oppositely to that in the $k_>$ state, the outgoing flux measured in the $k_<$ channel at $\theta=0$ will correspond to the wave vector $-k_<\hat{\b{x}}$. This is a spin-flipped state and will thus have zero contribution to the cross section. Hence, the orange lines in Fig.~\ref{fig:diffxsec} go to zero at $\theta=0$. Likewise, if the incident wave vector is $k_<\hat{\b{x}}$, then the spin-flipped states would be $-k_<\hat{\b{x}}$, detected at $\theta=0$, and $-k_>\hat{\b{x}}$, detected at $\theta=\pi$, corresponding to the zeroes of the differential cross section in those channels (red and green respectively in Fig.~\ref{fig:diffxsec}). 



\subsection{Total cross section}

\begin{figure}[t]
	\hspace*{-0.872in}\includegraphics[width=0.741\columnwidth]{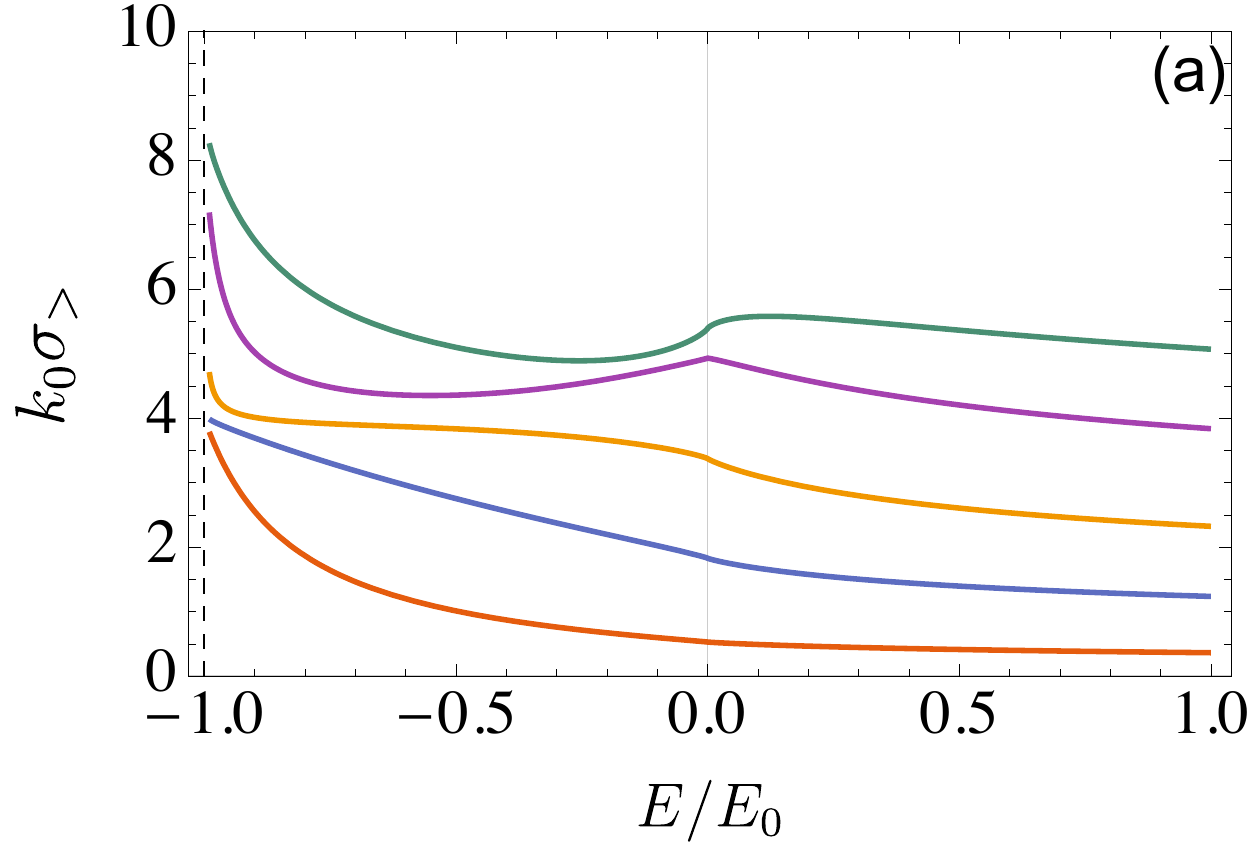}\\
	\hspace*{-0.872in}\includegraphics[width=0.741\columnwidth]{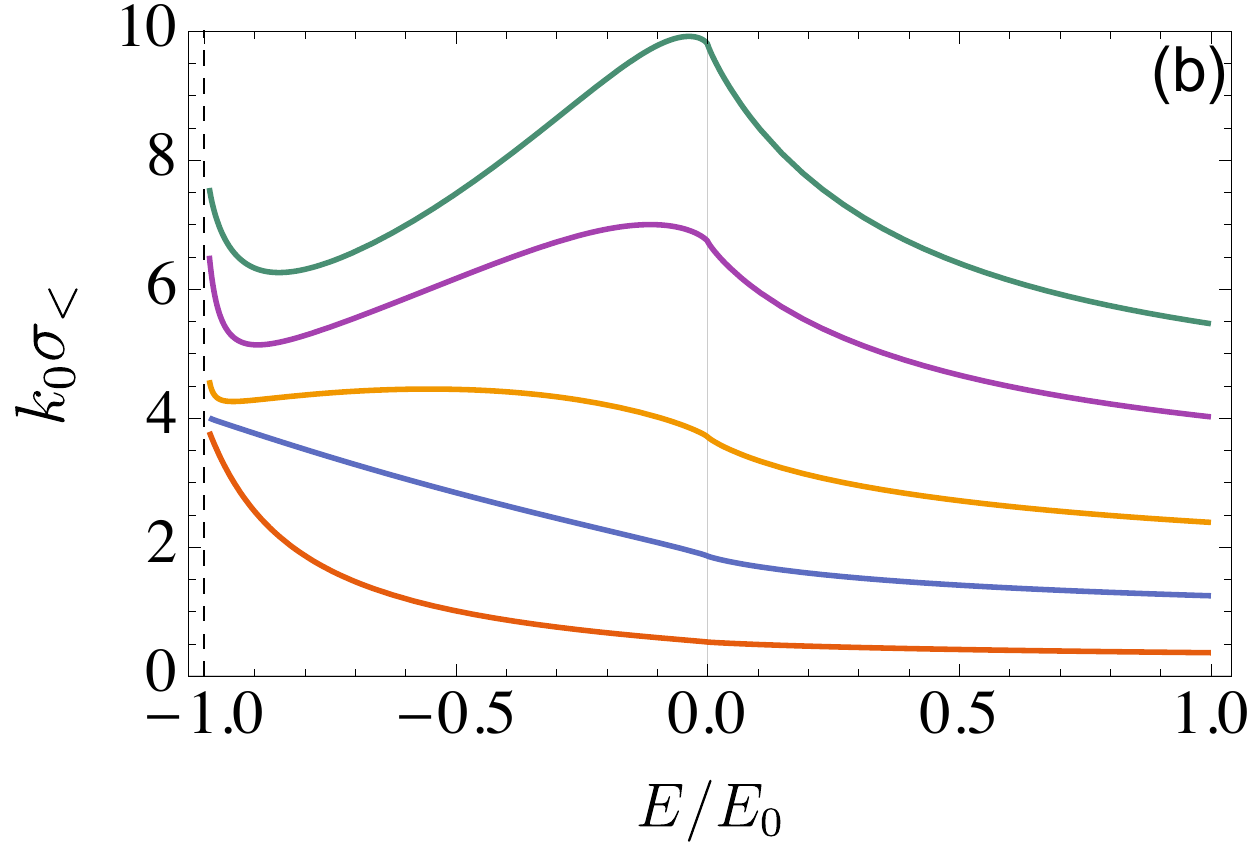}
	\includegraphics[width=1.02\columnwidth]{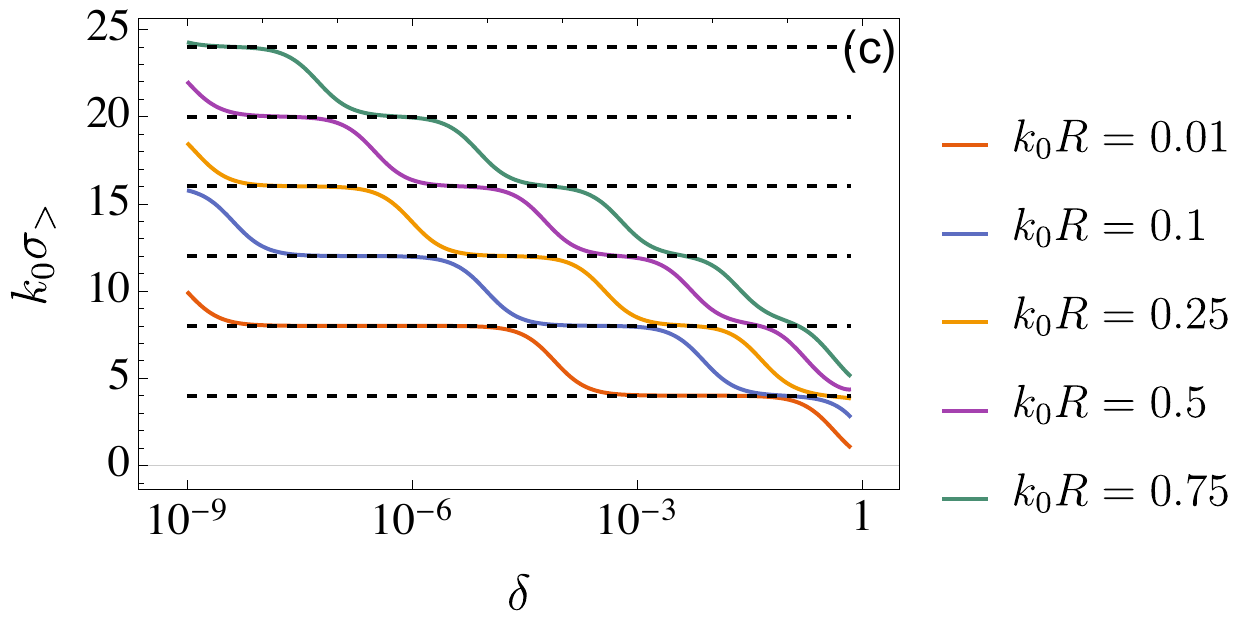}
\caption{Total cross section for various values of the dimensionless scatterer radius $k_0R$: (a) total cross section for incoming $k_>$ state as a function of energy, (b) total cross section for incoming $k_<$ state as a function of energy, (c) total cross section for incoming $k_>$ state as a function of $\delta$ on a log-linear scale. In each of (a) and (b), the cross section is also calculated in the $E>0$ regime. For $\sigma_<$ ($\sigma_>$), this shows scattering from an incident positive-helicity (negative-helicity) state. The vertical dashed line in (a) and (b) at $E/E_0=-1$ is a guide to the eye, showing the divergent behavior of all cross sections at the band bottom. The horizontal dashed lines in (c) show the plateaus at $k_0\sigma_>=4n$, $n=0,1,2,\ldots$}
\label{fig:totalxsecs}
\end{figure}

Integrating Eq.~\eqref{eq:diffxsec+} and \eqref{eq:diffxsec-} over $\theta$ gives the total cross sections $\sigma_\gtrless$ for an incident $k_\gtrless$ state,
\begin{eqnarray}
\sigma_>=\frac{2}{k_>}\sum_l\bigg[1-8\operatorname{Re}\bigg(c_l\frac{(-i)^l}{2\sqrt{2}}\bigg)\bigg],\label{eq:totalxsecplus}\\
\sigma_<=\frac{2}{k_<}\sum_l\bigg[1-8\operatorname{Re}\bigg(\tilde{b}_l\frac{(-i)^l}{2\sqrt{2}}\bigg)\bigg].\label{eq:totalxsecminus}
\end{eqnarray}
These are plotted in Fig.~\ref{fig:totalxsecs} as a function of the energy. For any value of the dimensionless radius of the scatterer $k_0R$, there is a singularity in the cross section at the band bottom $E\rightarrow-E_0$, due to the squared delta functions in Eq.~\eqref{eq:diffxsec_E0}. Equivalently, from Eq.~\eqref{eq:totalxsecplus} and \eqref{eq:totalxsecminus} we get the divergent sum $\sigma_\gtrless\rightarrow (2/k_0)\sum_l 1$ as $E\rightarrow-E_0$. Threshold singularities in the cross section are common to scattering problems in 2D (see Appendix~\ref{App:AppendixA}); however, in the conventional case without spin-orbit coupling such singularities are typically due to a prefactor of $1/k$ which diverges as $k\rightarrow 0$ at the bottom of a parabolic band~\cite{friedrich2013}. In the Rashba case, it is the sum over partial waves rather than the prefactor $1/k_0$ that diverges at the band bottom, since in that limit all $l$ channels contribute equally (Fig.~\ref{fig:S_amp2}).

In Fig.~\ref{fig:totalxsecs}(c), we zoom in on the region near the band bottom, and plot the total cross section $\sigma_>$ as a function of $\delta=\sqrt{1-|E|/E_0}$ on a log-linear scale. As the energy approaches the band bottom, the cross section increases in discrete steps and displays a series of plateaus that are increasingly flat as $\delta$ tends to zero on a logarithmic scale, with the onset of each plateau occurring at the threshold energy where a new $l$ channel contributes to the off-diagonal $S$-matrix elements [compare with Fig.~\ref{fig:S_amp2}(b)]. A similar behavior is found for $\sigma_<$. On these plateaus the total cross section is quantized in units of $4/k_0$,
\begin{align}\label{quantizedsigma}
\sigma_\gtrless=\frac{4n}{k_0},\,n=0,1,2,\ldots,
\end{align}
independently of the scatterer radius $R$. The way $\sigma_\gtrless$ approaches infinity as the energy nears the band bottom is thus much more complex than the smooth $1/k\propto 1/\sqrt{E}$ divergence (moderated by a logarithmic factor) found in the case without spin-orbit coupling where the $l=0$ partial wave ($s$-wave) dominates the low-energy behavior~\cite{friedrich2013}. An analogy with Landauer quantization of the conductance in 1D~\cite{landauer1957,vanwees1988,wharam1988} may lead one to conjecture that the quantization of the total cross section (\ref{quantizedsigma}) in the low-energy limit is a direct consequence of the emergent 1D behavior in that limit, observed in the extreme anisotropy of the differential cross section (\ref{eq:diffxsec_E0}).

\section{Delta-shell scattering}\label{sec:hardshell}

In the low-energy limit $E\rightarrow-E_0$, the $S$-matrix (\ref{UniversalSmatrix}) and, consequently, the differential cross section (\ref{eq:diffxsec_E0}) and plateau behavior of the total cross section (\ref{quantizedsigma}) were found to be completely independent of the range $R$ of the scattering potential. While this result suggests the form (\ref{UniversalSmatrix}) of the $S$-matrix is a universal feature of Rashba scattering in the low-energy limit, at least for spin-independent and rotationally invariant finite-range potentials $V(r)$, the possibility remains that Eq.~(\ref{UniversalSmatrix}) is a special feature of the hard-disk potential (\ref{Vharddisk}). To further support our conjecture of the universality of the low-energy $S$-matrix (\ref{UniversalSmatrix}), we consider the $E<0$ scattering problem for another scattering potential, the delta-shell potential:
\be
V(r)=V_0\delta(r-R).
\ee
Compared with the hard-disk potential (\ref{Vharddisk}), this potential has two tunable parameters, $V_0$ and $R$. In the region $r>R$, the wave function has the same form as Eq.~\eqref{eq:psi_generic}. For $r<R$, the Neumann functions $N_l(k_\gtrless r)$ must be eliminated for the solution to be regular at $r=0$. Thus,
\begin{eqnarray}
\Psi_{r<R}(r,\theta)&=&\sum_{l=-\infty}^\infty e^{il\theta}\bigg[a_l'\begin{pmatrix} J_l(k_>r)\\ -J_{l+1}(k_>r)e^{i\theta}\end{pmatrix}\nonumber\\
&&+b_l'\begin{pmatrix} J_l(k_<r)\\ -J_{l+1}(k_<r)e^{i\theta}\end{pmatrix}\bigg].
\end{eqnarray}
Consider an incident $k_>$ state. Then $a_l=0$, $d_l=\frac{i^l}{2\sqrt{2}}$, and there are four unknown coefficients. Continuity of the wave function at $r=R$ gives two equations,
\be\label{matching1}
\Psi_{r>R}(R,\theta)=\Psi_{r<R}(R,\theta),
\ee
and integrating the Schr\"odinger equation along the radial direction from $R-\epsilon$ to $R+\epsilon$ gives two more
\be\label{matching2}
\partial_r\Psi_{r>R}(R,\theta)-\partial_r\Psi_{r<R}(R,\theta)=2mV_0\Psi(R,\theta).
\ee
All four coefficients can thus be solved for, but their closed forms are too long to present here. Instead, we focus on the low-energy limit. At the band bottom, we have $k_<=k_>=k_0$ and the matching conditions (\ref{matching1})-(\ref{matching2}) may be written as the matrix equation
\be
M\begin{pmatrix}
a_l'\\
b_l'\\
b_l+\frac{i^l}{2\sqrt{2}}\\
c_l
\end{pmatrix}=\begin{pmatrix}
0\\
0\\
0\\
0
\end{pmatrix},
\ee
where $M$ is a $4\times4$ matrix containing Bessel and Hankel functions evaluated at $k_0R$. One can readily verify that $\det{M}\neq0$ for any nonzero value of $V_0$. Thus only the trivial solution $a_l'=b_l'=c_l=0$, $b_l=-\frac{i^l}{2\sqrt{2}}$ satisfies the matching conditions, which is precisely the result from hard-disk scattering. 

The $S$-matrix (\ref{UniversalSmatrix}) appears to be a universal feature of low-energy Rashba scattering in that it applies to both hard-disk and delta-shell potentials of any radius $R$ and magnitude $V_0$. We conjecture that this extends to any circularly symmetric, spin-independent potential of finite radius.


\section{Conclusion}

In summary, we have studied the scattering of electrons with Rashba spin-orbit coupling off spin-independent, circularly symmetric potentials in the negative-energy regime $E<0$, with a focus on the approach to the band bottom $E\rightarrow-E_0$. We find several features in this limit that appear to be insensitive to details of the scattering potential: the $S$-matrix approaches a purely off-diagonal form with both off-diagonal elements equal to negative one, and all angular momentum channels contribute equally at the band bottom; the differential cross section is increasingly peaked at forward and backward scattering angles; the total cross section increases by quantized steps as the energy approaches the band bottom. The quasi-1D character of these features supports and further expands Ref.~\cite{cappelluti2007}'s interpretation of reduction in effective dimensionality in the low-energy limit of Rashba systems. In the presence of harmonic potentials, the energy spectrum of Rashba systems is known to exhibit Landau-level-like quantization~\cite{Li2012}, which can be interpreted as yet another manifestation of dimensional reduction induced by spin-orbit coupling.

We conjecture the features we have found are universal, at least for spin-independent, circularly symmetric, finite-range potentials. It would be interesting to test this conjecture with other potentials in this class, and further see if it extends to spin-dependent but otherwise time-reversal-symmetric potentials. We expect some of the features we have discussed could be observed experimentally in low-density, strongly spin-orbit coupled 2D electron gases using scanning gate microscopy techniques, which have been used to image coherent electron flow~\cite{topinka2000,topinka2001}: concrete predictions to be compared directly with experiment such as simulated current maps could in principle be derived from the results presented in this work, for example by the method discussed in Ref.~\cite{walls2006}.

\acknowledgements

We thank F. Marsiglio for useful insights and discussion. J. H. was supported by NSERC. J.M. was supported by NSERC grant \#RGPIN-2014-4608, the Canada Research Chair Program (CRC), the Canadian Institute for Advanced Research (CIFAR), and the University of Alberta.


\appendix

\section{Spin-degenerate hard-disk scattering} \label{App:AppendixA}

\begin{figure}[t]
	\centering
	\includegraphics[width=\columnwidth]{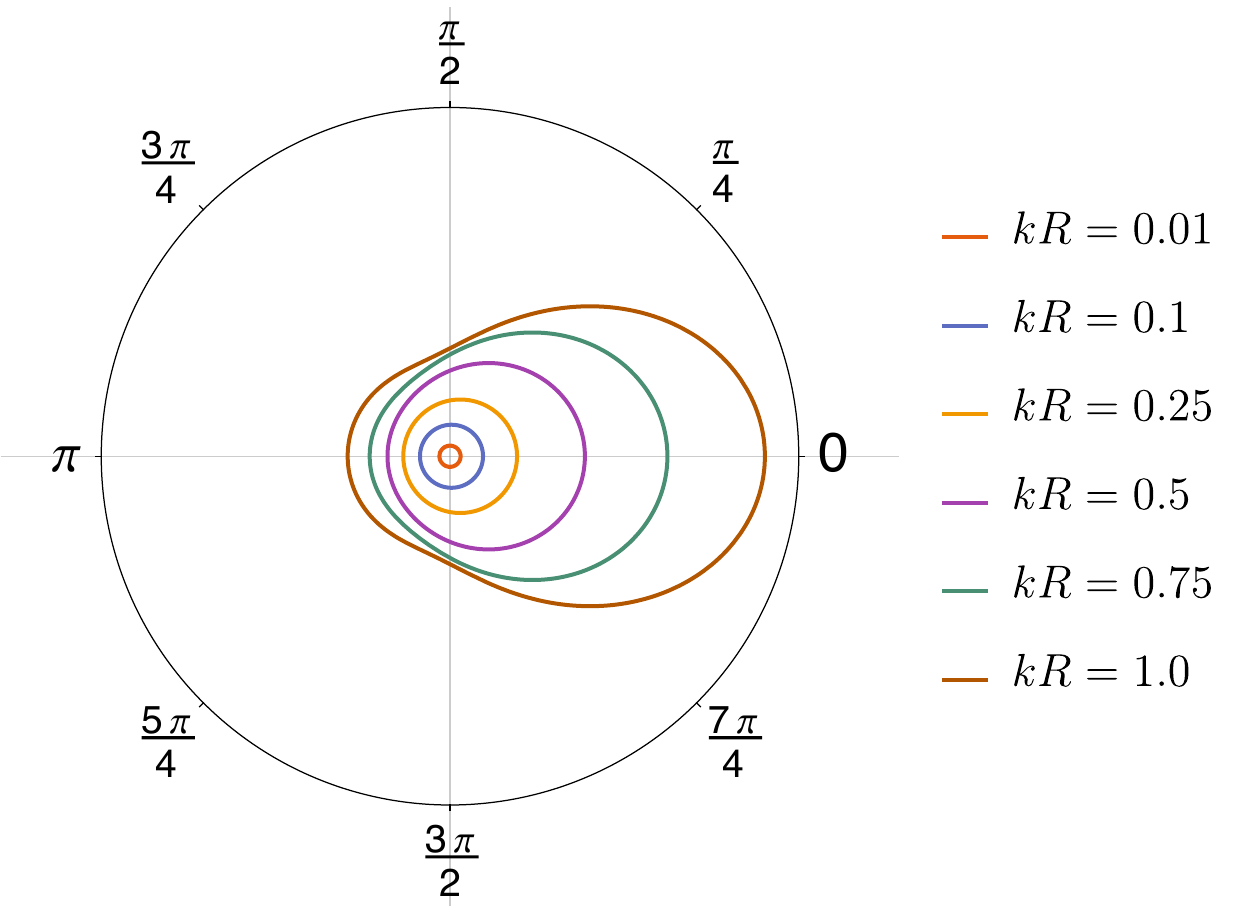} 
\caption{Polar plot of differential cross section for the spin-degenerate problem with various values of $kR$. The radius of each curve is the magnitude of $d\sigma/d\theta$ in units of $1/k$.}
\label{fig:spindegen}
\end{figure}

For comparison we present the results for electrons scattering off the hard-disk potential (\ref{Vharddisk}) in 2D but without spin-orbit coupling. In this case, the wave function in the scattering region $r>R$ is given by 
\be
\Psi(r,\theta)=\bigg(\frac{1}{\sqrt{2}}e^{ikx}+\sum_{l=-\infty}^{\infty}a_le^{il\theta}H_l^+(kr)\bigg)\eta,
\ee
where $\eta$ is an arbitrary spinor, and there is only a single wave vector $k=\sqrt{2mE}$ for each incident energy $E$. The matching condition \eqref{eq:match} gives two degenerate equations that determine the only unknown coefficient\
\be\label{SpinDegCoeff}
a_l=-\frac{i^l}{\sqrt{2}}\frac{J_l(kR)}{H_l^+(kR)}.
\ee
The incident and scattered current densities have magnitudes $|\vec{j}_{\rm in}|=\frac{k}{2m}$ and $|\vec{j}_{\rm sc}|=\frac{2}{\pi m r}|\sum_{l=-\infty}^\infty a_le^{i(\theta-\pi/2)l}|^2$ respectively. Equation \eqref{eq:diffxsec} then gives the differential cross section 
\be\label{eq:diffxsec1}
\frac{d\sigma}{d\theta}=\frac{4}{\pi k}\bigg|\sum_{l=-\infty}^\infty a_le^{i(\theta-\pi/2)l}\bigg|^2,
\ee
which is plotted in Fig.~\ref{fig:spindegen}. The cross section is isotropic in the long-wavelength limit, and forward scattering is enhanced as the wavelength is decreased.

In the long-wavelength limit, one may use the small-argument form of the Bessel functions,
\begin{eqnarray}
J_l(kr)&\approx&\frac{\epsilon_l}{|l|!}\bigg(\frac{kr}{2}\bigg)^{|l|},\\
N_l(kr)&\approx&\begin{cases} 
      -\displaystyle\frac{\epsilon_l(|l|-1)!}{\pi}\left(\frac{2}{kr}\right)^{|l|}, & l\ne0, \\
     \displaystyle\frac{2}{\pi}\left[\ln\left(\frac{kr}{2}\right)+\gamma\right], & l=0,\\
   \end{cases}
\end{eqnarray}
where $\gamma$ is Euler's constant and
\be
\epsilon_l=\begin{cases} 
      1, & l>0, \\
     (-1)^l, & l<0. \\
   \end{cases}
   \ee
In this limit, the coefficient (\ref{SpinDegCoeff}) is
 \be\label{eq:al}
 a_l\approx\begin{cases}
 -\displaystyle\frac{i^{l}}{\sqrt{2}}\left[1-\frac{i}{\pi}(|l|-1)!|l|!\left(\frac{kR}{2}\right)^{2|l|}\right]^{-1}, & l\ne 0,\\
-\displaystyle\frac{1}{\sqrt{2}}\left\{1+i\frac{2}{\pi}\left[\ln\left(\frac{kR}{2}\right)+\gamma\right]\right\}^{-1}, & l=0. \\
 \end{cases}
 \ee
 
 It is more common to write scattering quantities in terms of the phase shift $\delta_l$ (which is more ambiguous in the case of multiple scattering channels)~\cite{friedrich2013}. In this case the differential cross section (\ref{eq:diffxsec1}) is written as
 \be
 \frac{d\sigma}{d\theta}=\frac{2}{\pi k}\bigg|\sum_{l=-\infty}^\infty\sin\delta_le^{i(l\theta+\delta_l)}\bigg|^2.
 \ee
 Comparison with Eq.~\eqref{eq:diffxsec1} and \eqref{eq:al} gives the phase shifts
 \be
 \cot\delta_l=\begin{cases}
 -\displaystyle\frac{1}{\pi}(|l|-1)!|l|!\left(\frac{2}{kR}\right)^{2|l|}, & l\ne 0,\\
 \displaystyle\frac{2}{\pi}\left[\ln\left(\frac{kR}{2}\right)+\gamma\right], & l=0.
 \end{cases}
 \ee
 Note that even in the long-wavelength limit, the differential cross section and phase shift retain a dependence on $R$ in contrast to the case with spin-orbit coupling. However, the singularity in the cross section at threshold is a common feature of scattering in 2D~\cite{friedrich2013}.
 
\section{Symmetry of the $S$-matrix}\label{app:symmetry}

Here we show that the symmetry of the $S$-matrix $S^l=(S^l)^T$ for each angular momentum component $l$ is a consequence of the combination of two symmetries: time-reversal symmetry, and a symmetry under reflection about the $x$ axis, i.e., symmetry under $y\rightarrow-y$.
 
The action of the time-reversal operator $T$ on an arbitrary spinor $\psi(\b{r})=\psi_\uparrow(\b{r})\lvert\uparrow\rangle+\psi_\downarrow(\b{r})\lvert\downarrow\rangle$ (with $\lvert\uparrow\rangle=(1,0)$ and $\lvert\downarrow\rangle=(0,1)$ the eigenvectors of $\sigma_z$) is given by
\begin{align}
T\psi(\b{r})=\psi_\uparrow^*(\b{r})\lvert\downarrow\rangle-\psi_\downarrow^*(\b{r})\lvert\uparrow\rangle=-i\sigma_y\psi^*(\b{r}).
\end{align}
One can check by explicit calculation that the Hamiltonian (\ref{Hpolar}) obeys the relation
\begin{align}
\sigma_y H(r,\theta)\sigma_y=H^*(r,\theta),
\end{align}
which is a statement of time-reversal symmetry. Thus if $\psi_E(r,\theta)$ is an eigenstate of $H(r,\theta)$ with energy $E$, the state $T\psi_E(r,\theta)=-i\sigma_y\psi^*_E(r,\theta)$ is also an eigenstate of $H(r,\theta)$ at the same energy. Likewise, the Hamiltonian obeys the relation
\begin{align}
\sigma_y H(r,\theta)\sigma_y=H(r,-\theta),
\end{align}
which is a statement of reflection symmetry about the $x$ axis (i.e., $y\rightarrow-y$ or $\theta\rightarrow-\theta$). Indeed, because the incident plane wave propagates in the $x$ direction and the scattering potential is rotationally symmetric, this is a symmetry of the scattering geometry (Fig.~\ref{fig:scatter}). If $\psi_E(r,\theta)$ is an eigenstate of $H(r,\theta)$ with energy $E$, the state $\sigma_y\psi_E(r,-\theta)$ is also an eigenstate of $H(r,\theta)$ at the same energy~\cite{Yeh2006}. Combining these two symmetries, we find that $\psi_E^*(r,-\theta)$ is an eigenstate of $H(r,\theta)$ with energy $E$ if $\psi_E(r,\theta)$ is.

We can use the fact we have just derived to constrain the form of the $S$-matrix. Because the scattering states (\ref{eq:wavefunck+}) and (\ref{eq:wavefunck-}) described by the $S$-matrix $S^l$ are eigenstates of the Hamiltonian with energy $E$, the states $\psi_>^*(r,-\theta)$ and $\psi_<^*(r,-\theta)$ are also eigenstates of the Hamiltonian with the same energy, and should thus be described by the same $S$-matrix. We first introduce the notation
\begin{eqnarray}
\phi^\text{in}_\gtrless(r,\theta)&=&\sqrt{k_\gtrless}\left(\begin{array}{c}
H_l^\mp(k_\gtrless r) \\
-H_{l+1}^\mp(k_\gtrless r)e^{i\theta}
\end{array}\right),\\
\phi^\text{out}_\gtrless(r,\theta)&=&\sqrt{k_\gtrless}\left(\begin{array}{c}
H_l^\pm(k_\gtrless r) \\
-H_{l+1}^\pm(k_\gtrless r)e^{i\theta}
\end{array}\right).
\end{eqnarray}
Because the combined action of complex conjugation and reversing the sign of $\theta$ leaves the angular factor $e^{il\theta}$ invariant, we can consider one $l$ component at a time. Ignoring a constant multiplicative factor, for a given $l$ and in the asymptotic region $k_\gtrless r\gg 1$ one has
\begin{align}
\psi_>(r,\theta)&\sim\phi^\text{in}_> + S_{>>}^l\phi^\text{out}_> + S_{><}^l\phi^\text{out}_<,\label{scatt1}\\
\psi_<(r,\theta)&\sim\phi^\text{in}_< + S_{<>}^l\phi^\text{out}_> + S_{<<}^l\phi^\text{out}_<.\label{scatt2}
\end{align}
The combined action of complex conjugation and reversing the sign of $\theta$ interchanges incoming and outgoing circular waves,
\begin{align}
\phi_\gtrless^\text{in}(r,-\theta)^*=\phi_\gtrless^\text{out}(r,\theta),
\end{align}
such that for a given $l$ one has
\begin{align}
\psi_>^*(r,-\theta)&\sim\phi^\text{out}_> + (S_{>>}^l)^*\phi^\text{in}_> + (S_{><}^l)^*\phi^\text{in}_<,\label{symm1}\\
\psi_<^*(r,-\theta)&\sim\phi^\text{out}_< + (S_{<>}^l)^*\phi^\text{in}_> + (S_{<<}^l)^*\phi^\text{in}_<.\label{symm2}
\end{align}
Because the scattering states (\ref{symm1}) and (\ref{symm2}) are degenerate, an arbitrary linear superposition of those two states is also a valid scattering state at the same energy. In particular, we can construct linear superpositions $\tilde{\psi}_>(r,\theta)$ and $\tilde{\psi}_<(r,\theta)$ that take the standard form (\ref{scatt1})-(\ref{scatt2}) of an incoming circular wave $\phi^\text{in}_\gtrless$ plus outgoing circular waves $\phi^\text{out}_\gtrless$ multiplied by appropriate coefficients,
\begin{align}
\tilde{\psi}_>(r,\theta)\sim\phi^\text{in}_> +\left(\frac{S_{<<}^l}{\det S^l}\right)^*\phi^\text{out}_> - \left(\frac{S_{><}^l}{\det S^l}\right)^*
 \phi^\text{out}_<,\\
 \tilde{\psi}_<(r,\theta)\sim\phi^\text{in}_< -\left(\frac{S_{<>}^l}{\det S^l}\right)^*\phi^\text{out}_> + \left(\frac{S_{>>}^l}{\det S^l}\right)^*
 \phi^\text{out}_<.
\end{align}
Comparing with Eq.~(\ref{scatt1})-(\ref{scatt2}), we obtain the relations
\begin{align}\label{relation}
S_{>>}^l&=\left(\frac{S_{<<}^l}{\det S^l}\right)^*, \hspace{5mm}
S_{><}^l=- \left(\frac{S_{><}^l}{\det S^l}\right)^*, \nonumber\\
S_{<>}^l&=-\left(\frac{S_{<>}^l}{\det S^l}\right)^*, \hspace{5mm}
S_{<<}^l=\left(\frac{S_{>>}^l}{\det S^l}\right)^*.
\end{align}
Using the inverse of the $S$-matrix
\begin{align}
(S^l)^{-1}=\frac{1}{\det S^l}\left(\begin{array}{cc}
S_{<<}^l & -S_{><}^l \\
-S_{<>}^l & S_{>>}^l
\end{array}\right),
\end{align}
as well as its unitarity $(S^l)^{-1}_{\alpha\beta}=S^*_{\beta\alpha}$, the first and fourth relations in (\ref{relation}) are trivial and the second and third give
\begin{align}
S_{><}^l=S_{<>}^l,
\end{align}
i.e., $S^l=(S^l)^T$.
 
\bibliography{rashba1p}

\end{document}